\documentstyle[cf98]{article}

\input{psfig}

\def\be{\begin{equation}}
\def\ee{\end{equation}}
\def\bea{\begin{eqnarray}}
\def\eea{\end{eqnarray}}

\def\ben{\begin{eqnarray}}
\def\enn{\end{eqnarray}}
\def\ov{\over\displaystyle\strut}
\def\dst{\displaystyle\phantom{\mid}}
\def\l({\left(}
\def\r){\right)}

\begin{document}

\title{TESTING BUDA-LUND HYDRO MODEL ON PARTICLE CORRELATIONS AND 
SPECTRA IN NA44, WA93 AND WA98 HEAVY ION EXPERIMENTS}

\author{\underline{A. STER}}

\address{MTA KFKI Research Institute for Technical Physics and Materials Science, 
\\H-1525 Budapest 114, POB 49, Hungary
\\for the WA93 and WA98 collaborations 
\\E-mail: ster@rmki.kfki.hu} 

\author{T. CS{\"O}RG\H O}

\address{MTA KFKI Research Institute for Particle and Nuclear Physics, 
\\H-1525 Budapest 114, POB 49, Hungary \\E-mail: csorgo@sunserv.kfki.hu} 

\author{B. L{\"O}RSTAD}

\address{Physics Department, Lund University,\\
S-221 00 Lund, POB 118, Sweden
\\for the NA44 collaboration
\\E-mail: bengt@quark.lu.se}


\maketitle\abstracts{ 
        Analytic and numerical approximations
        to a hydrodynamical model 
        describing longitudinally expanding,
        cylindrically symmetric, finite systems are fitted to preliminary
	NA44 data 
        measured in 200 AGeV central $S + Pb$ reactions.
        The model describes the measured spectra 
        and HBT radii of pions, kaons and protons, simultaneously.
        The source is characterized by 
        a central freeze-out temperature of $T_0 = 154 
        \pm 8 \pm 11 $ MeV, a ``surface" temperature of  
        $T_r = 107 \pm 28 \pm 18$ MeV and by  
        a well-developed transverse flow, 
        $\langle u_t \rangle = 0.53 \pm 0.17 \pm 0.11$. 
        The transverse geometrical radius  and the
        mean freeze-out time are found to be  
        $R_G = 5.4 \pm 0.9 \pm 0.7$ fm and  
        $\tau_0 =  5.1 \pm 0.3 \pm 0.3$ fm/c, respectively.
        Fits to preliminary WA93 200 AGeV S + Au and WA98 158 AGeV
	Pb + Pb data dominated by pions indicate similar model parameters.
        The {\it absolute normalization} of  
        the  measured particle spectra together with 
        the experimental determination of both 
        the statistical and the systematic errors
        were needed to obtain successful fits. 
}

\section{Introduction}
	The reconstruction of the space-time picture of particle 
	emission in high energy heavy ion and particle physics 
	became a focal point of current research interest.
	In high energy heavy ion physics, the space-time information
	on particle production is needed to determine the volume 
	of hot hadronic matter at freeze-out~\cite{qm} while in particle physics
	Bose-Einstein correlations of pions from different
	decaying $W$ mesons may result in large contribution to the 
	systematic errors of $W$-mass measurements 
	at LEP II~\cite{leif}, a major problem 
	in forthcoming precision determination of $W$ mass.

	In this paper, we attempt to reconstruct the
	space-time picture of high energy heavy ion reactions
	by checking whether the hydrodynamical model 
	of ref.~\cite{3d} (referred to as the Buda-Lund hydro 
	parameterization) is able to fit NA44 data on
	two-particle correlations and single-particle spectra 
	at S + Pb 200 AGeV central reactions at CERN SPS. 
	Data from WA93 200 AGeV S + Au and WA98 158 AGeV
        Pb + Pb experiments are also used to confirm the
	reliability of the fit results.
	The  Buda-Lund hydro parameterization~\cite{3d} 
	corresponds to a  class of
	longitudinally expanding, cylindrically symmetric, finite
	systems, where the  local rest density 	distribution,
	the local inverse temperature distribution and the 
	freeze-out proper-time distribution is characterized 
	by their means and  variances, respectively. 
 	A scaling longitudinal flow field
	is assumed together with a linear transverse flow profile.
	As a consequence of the cylindrical symmetry of the model, 
	a $1/\sqrt{m_t}$ scaling of the HBT radius parameters
	was predicted in certain limited regions 
	of the parameter space~\cite{3d,3dqm}. 
	Such a scaling is difficult to obtain in other type of models,
	however, the scaling law is satisfied by not only the NA44 data
	on S+Pb reactions~\cite{na44mt} but also by the NA22 data
	on hadron-proton interactions at CERN SPS~\cite{na22mt}
	and, to some extent, by preliminary $e^+ + e^-$ annihilation data
	in two-jet events at LEP-I ~\cite{DELPHImt}. 
	In ref.~\cite{1d,3d,3dqm} it was observed for the first time,
	that the source parameters can be determined precisely
	only if {\it  a simultaneous analysis of the particle spectra and
	correlation functions} is performed. 
	Such results on data fitting to analytical and numerical
	approximations are presented in the body of this paper.

\section{The model and its re-parameterization}

\subsection{The model}
	
	The Buda-Lund hydroparameterization~\cite{3d} makes 
	difference between
        the central (core) and the outskirts (halo) 
	regions of the collided
        matter. The following emission function $S_c(x,p)$ applies to a
	hydrodynamically evolving three dimensionally expanding, 
	cylindrically symmetric finite system:

\ben
        S_{c}(x,p) \, d^4 x & = & {\dst  g \ov (2 \pi)^3} \,
        {\dst d^4 \Sigma^{\mu}(x) p_{\mu} \ov
        \exp\l({\dst  u^{\mu}(x)p_{\mu} \ov  T(x)} -
        {\dst \mu(x) \ov  T(x)}\r) - 1} \ , \label{e:s}
\enn
where the subscript $_c$ refers to the core,
the factor $ d^4 \Sigma^{\mu}(x) p_{\mu} $ 
describes the flux of particles through a
finite, narrow layer of freeze-out hypersurfaces.
It is assumed that any of the
layers can be labelled  by a unique value of
$ \tau = \sqrt{t^2 - z^2} $, and the random variable $\tau$
is characterized by a probability distribution, such that
\ben
         d^4 \Sigma^{\mu}(x) p_{\mu} & = &
         m_t \cosh[\eta - y]  \, H(\tau) d\tau \, \tau_0 d\eta \, dr_x \, dr_y \
 ,
\enn
	Here $m_t = \sqrt{m^2 + p_x^2 + p_y^2}$ stands for the transverse
	mass, the rapidity $y$ and the space-time rapidity $\eta$ are 
	defined as $y = 0.5 \log\left[(E+ p_z) / (E - p_z)\right]$ and
	$\eta = 0.5 \log\left[(t+ z) / (t - z)\right]$ and
        the  duration of particle emission is characterized by
        $H(\tau) \propto \exp(-(\tau-\tau_0)^2
        /(2 \Delta\tau^2))$.  Here
        $\tau_0$ is the mean emission time,
        $\Delta \tau$ is the duration of the emission in (proper) time.
        The four-velocity  and the local temperature and density profile
         of the expanding matter is given by
\ben
        u^{\mu}(x) & = & \l( \cosh[\eta] \cosh[\eta_t],
        \, \sinh[\eta_t] {\dst r_x \ov r_t},
        \, \sinh[\eta_t] {\dst r_y \ov r_t},
        \, \sinh[\eta] \cosh[\eta_t] \r), \\
	\sinh[\eta_t] & = & b {r_t \ov \tau_0}, 
	\qquad r_t \,\,  = \,\, \sqrt{r_x^2 + r_y^2}.
\enn
	The inverse temperature profile is characterized by the 
	central value and its variance in transverse and temporal
	direction, and we assume a Gaussian shape of the local density
	distribution:
\ben
{\dst 1 \ov T(x)} & =  &
	{\dst 1 \ov T_0 } \,\,
	\left( 1 + a^2 \, {\dst  r_t^2 \ov 2 \tau_0^2} \right) \,
	\left( 1 + d^2 \, {\dst (\tau - \tau_0)^2 \ov 2 \tau_0^2  } 
	\right) ,
	\\
	{\dst \mu(x) \ov T(x) }  & = &  {\dst \mu_0 \ov T_0} -
        { \dst r_x^2 + r_y^2 \ov 2 R_G^2}
        -{ \dst (\eta - y_0)^2 \ov 2 \Delta \eta^2 }, \label{e:mu}
\enn
	where $\mu(x)$ is the chemical potential and $T(x)$ is the local
        temperature characterizing the particle emission. Note that the
	strength of the transverse changes of the temperature profile,
	the gradient of the transverse flow and the 
	strength of the temporal  changes of the temperature
	profile are controlled by the dimensionless parameters
	$a$, $b$ and $d$, respectively.

\subsection{Analytic approximations}

	In Ref.~\cite{3d}, the Boltzmann
	approximation to the above emission function 
	was evaluated in an analytical manner, 
	applying approximations around the saddle point of the
        emission function. The resulting formulas express the
        Invariant Momentum Distribution (IMD) and the Bose-Einstein
	correlation function (BECF)  in  an analytic way.
	The resulting analytical formulas are given in
	refs.~\cite{3d,3dqm,3dnum,na22} and shall not be recapitulated herewith.
	This Boltzmann approximation was also applied 
	to the numerical approximate evaluation of the model,
	as given in the next section.

\subsection{Core/halo correction}
	
        The effective intercept parameter 	
	$\lambda_*(y,m_t)$ of the Bose-Einstein correlation function 
	controls the core ratio in the 
        particle production in the core/halo picture developed in
	refs.~\cite{chalom,chalo,nhalo}. 
	With this factor the total invariant spectrum
	in $y$ rapidity and transverse mass $m_t$ follows as

\ben
        {\dst d^2 n\ov dy \, dm_t^2 } & = &
                        {\dst 1 \ov \sqrt{\lambda_*} }
                        {\dst d^2 n_{c} \ov dy \, dm_t^2 } \,\, =  \,\,
		{\dst \pi \ov  \sqrt{\lambda_*} } \,
                {\int S_c(x,p) \, d^4 x}
, 
\enn
	The momentum dependence of $\lambda_*$ parameter was mesured by NA44
	in ref.~\cite{na44mt}, although with a very limited momentum
	resolution.

\subsection{Re-parameterization}

	We introduce the surface temperature $T_r = T (r_x = r_y = R_G, 
	\tau = \tau_0)$ and the temperature after most of the particles
	were emitted as $T_t = T(r_x = r_y = 0; 
	\tau = \tau_0 + \sqrt{2} \Delta\tau)$. Here $R_G$ stands for 
	the transverse geometrical radius of the source, $\tau_0$ denotes
	the mean freeze-out time, $\Delta \tau$ is the duration of the
	particle emission and we denote the temperature field by $T(x)$.
	The central temperature at mean freeze-out time is denoted by
	$T_0 = T(r_x = r_y = 0; \tau = \tau_0)$.

	Then the relative transverse and temporal temperature decrease
	can be introduced as
\ben
	\langle {\Delta T \over T}\rangle_r & = & {T_0 - T_r \ov T_r}, 
 \qquad \langle {\Delta T \over T}\rangle_t  =  {T_0 - T_t \ov T_t} 
\enn
	and it is worthwhile to introduce the mean transverse flow
	as the transverse flow at the geometrical radius as 
\be
 	\langle u_t \rangle = b \, {R_G \over \tau_0}.
\ee
	Hence the 3 dimensionless parameters can
	be re-expressed with the physical parameters introduced
	above as
\ben
         {a^2} & = & {\tau_0^2 \over R_G^2} \,\,
			\langle {\Delta T\over T}\rangle_r,  
\qquad   {b }  =  { \tau_0 \over R_G }\,\, \langle u_t \rangle,  
\qquad   {d^2 }  =  {\tau_0^2 \over \Delta \tau^2} \,\,
			\langle {\Delta T\over T}\rangle_t.
\enn 

\section{Fitting the model to data}

	The model has been tested in two different ways using the
        analytical approximation referenced above and the numerical
	approximation
        that is based on a numerical integration of the emission
        function. The kinematic parameters are
        fitted simultaneously to preliminary IMD and HBT radii measured
        by the CERN NA44 experiment in central $S + Pb$ 
	collisions at 200 AGeV. 

\begin{table}[t]
\caption{Parameters from simultaneous fitting of preliminary
particle spectra and HBT
radius parameters with analytic and numeric approximations to the
hydrodynamical core model. The table shows the combined results of the two
sorts of fits for the NA44 S + Pb and the analytic fit
results for the WA93 S + Au and the WA98 Pb + Pb reactions.
The errors on WA93 and WA98 parameters are preliminary, see Section 3.} 
\vspace{0.2cm}
\begin{center}
\footnotesize
\begin{tabular}{|l|l|l|l|l|l|l|}
\hline
\null		& \multicolumn{2}{|c|}{NA44 S+Pb} &
		  \multicolumn{2}{|c|}{WA93 S+Au} &
		  \multicolumn{2}{|c|}{WA98 Pb+Pb} \\ 
\hline
	Parameter    & Value & Errors (Stat \&  Syst.)
		     & Value & Errors  
		     & Value & Errors \\
\hline
	$T_0$  [MeV] 		& 154   & $\pm$ 8     $\pm$ 11  
				& 154   & $\pm$ 8                    
				& 146   & $\pm$ 3                         \\
	$\tau_0$ [fm/c]    	& 5.1   & $\pm$ 0.3   $\pm$ 0.3  
				& 4.7   & $\pm$ 0.5    
                                & 5.0   & $\pm$ 0.2                       \\
	$R_G$   [fm]   		& 5.4   & $\pm$ 0.9   $\pm$ 0.7   
                                & 4.2   & $\pm$ 0.5
	                        & 7.1   & $\pm$ 0.3                       \\
	$\Delta\eta$ 		& 1.6   & $\pm$ 0.3   $\pm$ 0.3   
                                & 1.6   & $\pm$ 0.7
	                        & 1.6   & $\pm$ 0.3                       \\
	$\Delta\tau$ [fm/c]	& 0.3   & $\pm$ 0.3   $\pm$ 0.3  
                                & 0.9   & $\pm$ 0.8
                                & 1.7   & $\pm$ 0.1                        \\
	a                       & 0.63  & $\pm$ 0.09  $\pm$ 0.01       
                                & 0.21  & $\pm$ 0.11
                                & 0.06  & $\pm$ 0.06                        \\
	b                       & 0.50  & $\pm$ 0.06  $\pm$ 0.06 
                                & 0.73  & $\pm$ 0.18
                                & 0.36  & $\pm$ 0.04                        \\
	d                       & 4.9   & $\pm$ 1.8   $\pm$ 1.1    
                                & 4.7   & $\pm$ 4.3
                                & 8.1   & $\pm$ 0.6                        \\
\hline
\end{tabular}
\end{center}

\caption{Parameters calculated from the results in Table 1.
Preliminary fits for NA44 using directly these parameters indicate
similar values with smaller errors.}
\vspace{0.2cm}
\begin{center}
\footnotesize
\begin{tabular}{|l|l|l|l|}
\hline
	Calculated parameters 
		     & Value  $\pm$ Error 
		     & Value  $\pm$ Error 
		     & Value  $\pm$ Error \\
\hline
	$\langle {\Delta T \over T}\rangle_r$ & 0.44 $\pm$ 0.45 $\pm$ 0.22   
		& 0.04  $\pm$ 0.09	& 0.01  $\pm$ 0.04 \\ 
	$\langle {\Delta T \over T}\rangle_t$ & 0.08 $\pm$ 0.62 $\pm$ 0.56 
		& 0.81  $\pm$ 13.30              & 7.50  $\pm$ 3.14 \\
	$T_r$ [MeV] 		& 107 $\pm$ 28 $\pm$ 18
		& 148  $\pm$ 19                  & 145  $\pm$ 8 \\
	$T_t$ [MeV] 		& 143 $\pm$ 61 $\pm$ 56 
		& 85  $\pm$ 77                  & 17  $\pm$ 10 \\
	$\langle u_t \rangle$ & 0.53 $\pm$ 0.17 $\pm$ 0.11
		& 0.66  $\pm$ 0.34              & 0.51  $\pm$ 0.10 \\
\hline
\end{tabular}
\end{center}
\vspace{-0.2cm}
\end{table}
	
	As emphasized previously, absolutely normalized data were 
	utilized in these fits, which were 
	performed with the help of the
	CERN function minimization package MINUIT.
	Moreover, core/halo correction $\propto$ 
	${1 / \sqrt{\lambda_*}}$ is applied
	and the corresponding errors are propagated properly.
	Due to the these conditions a unique minimum is found,
	and the strongly coupled, normalization sensitive
	$d$ and $\Delta \tau$ parameters are determined. 
	In contrast, if the data are fitted without 
	absolute normalization, we reproduce the results 
	in ref.~\cite{3ds96} and we obtain big errors 
	on these parameters.



	In the numeric approximation scheme we evaluate the
	means and the variances of the core as suggested  in ref.~\cite{uli}.
        Since this numeric approximation scheme is not an exact calculation, 
	but an approximation in a different way than the analytic approach,
	we use it to estimate the systematic errors of the model
	parameters.

	The analytic and numeric results are combined in Table 1 to 
	estimate the model parameters and their errors properly.
	On Figure 1, the analytic and numeric fits to measured data
	are shown simultaneously.
        The parameters $a$, $b$ and $d$ are transformed
        to the corresponding relative transverse temperature decrease,
        the mean transverse flow and the relative temporal temperature
        decrease on Table 2.

        A comparision between the numerical and  the analytical
        approximation schemes
        indicates that the minima found by the two rather different
        fitting methods coincide within 2 standard deviations
        for each parameter of the model.
        To estimate the systematic errors
        half of the difference between
        the minima of the two fits is evaluated
        for each parameter.
        To estimate the best values of the fit parameters,
        the mean of the analytic and numerical minima is taken.
        To estimate the errors on these values,
         the statistical error equals  to the bigger of
        statistical error of the numerical and  the analytical fits,
        the systematical error is defined as above.

	Fits to preliminary data of the WA93 and WA98 experiments
	seem to provide similar 
	source parameter values like those obtained by NA44. However,
	some of the parameters have big errors because
	the particles are unidentified in these experiments and
	statistics allowed for determination of
 	only one HBT radius in each ($side, out, long$) directions in the
	Bertsch-Pratt frame. The normalization of the WA98 spectrum
	was fixed manually, which resulted in artificially small errors 
	in Table 1.
	As an indication, the following characteristic values 
	are obtained from fits using the analytic model approximation. 
	In case of WA93 we get a central freeze-out temperature of $T_0 = 154
        \pm 8$ MeV and a mean transverse flow of $\langle u_t \rangle =
	0.66 \pm 0.34$. Analysis of WA98 data shows that the corresponding
	parameters are $146 \pm 3$ MeV and $0.51 \pm 0.10$, respectively.

\begin{figure}[t]
\begin{center}
\vskip -1.0cm
\hskip 4.0cm
\psfig{figure=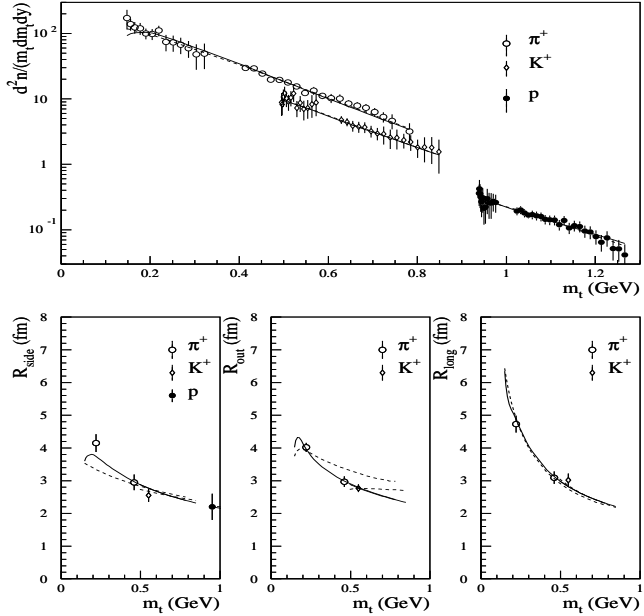,width=9.5cm,height=10.5cm}
\end{center}
\vskip -1.0cm
\caption{ Simultaneous fitting of particle spectra and HBT radius parameters.
Full line shows the fit using the analytical approximation, dashed line shows
the result of the numerical approximation.
Data points on the
spectra represent the $y$-averaged NA44 2 dimensional 
$d^2 n \ov m_t \, dm_t \, dy$ distribution. HBT data are found
in ref. $^5$.
} 
\end{figure}

\section{Discussion}
	
\subsection{The Source of particles in space-time}

\begin{figure}[t]
\begin{center}
\vskip -1.0cm
\hskip 3.5cm
\psfig{figure=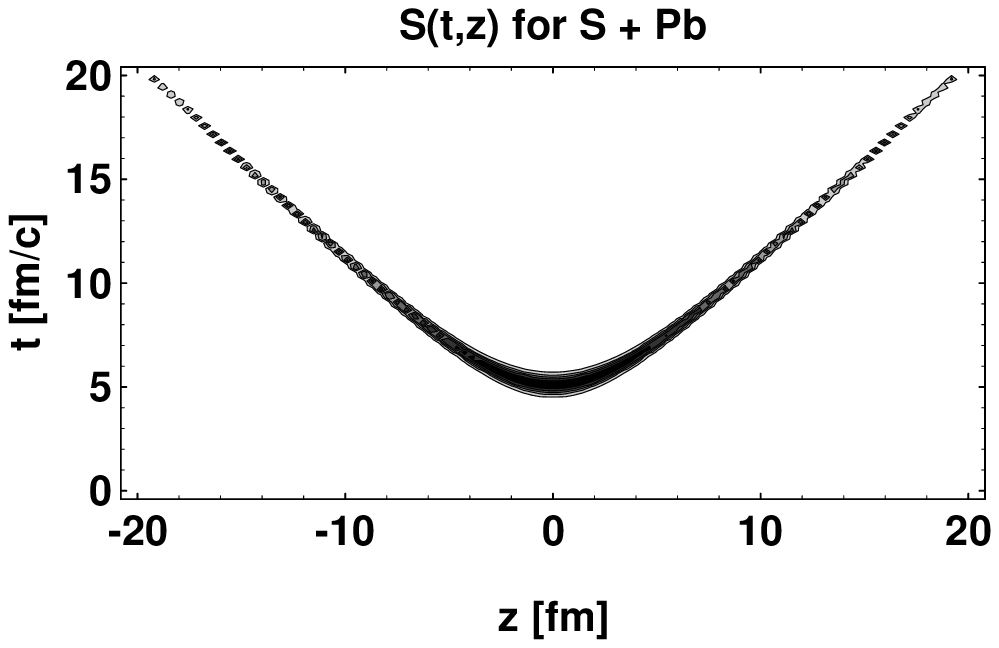,width=9.0cm,height=6.5cm}
\end{center}
\vskip -1.5cm
\caption{
Contour plot of the reconstructed source function for $S + Pb$ reactions
as measured by the NA44 collaboration. Note the large value of the
mean freeze-out time and the narrow width of the emission zone in
proper-time.  
}

\begin{center}
\vskip -1.0cm
\hskip 3.5cm
\psfig{figure=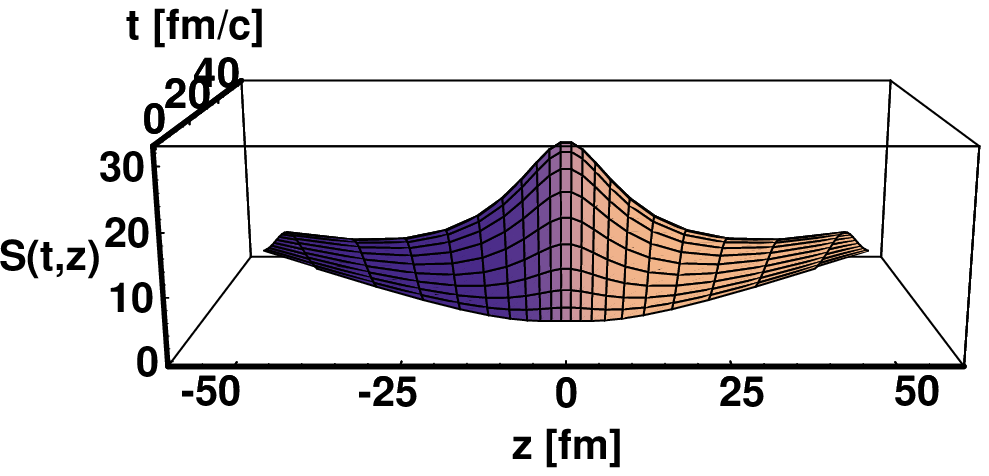,width=9.0cm,height=6.5cm}
\end{center}
\vskip -1.5cm
\caption{
Parametric plot of the reconstructed source function for $S + Pb$ reactions
as measured by the NA44 collaboration. The source is the same as that
of Figure 2, but a different view-point is chosen to show the shape of
emission probabilities for large values of $| z|$ in the mid-rapidity frame.
}
\end{figure}

	On Figures 2 and 3 we indicate the reconstructed space-time
	distribution of the source of particles in 200 AGeV central
	$S + Pb $ reactions, as a function of the time variable $t$
	and the coordinate along the beam direction, $z$, both  measured
	in the mid-rapidity frame that moves in the laboratory with 
	$y_0 = 3.0$. The momentum-integrated
	emission function at $r_x = r_y = 0$ is given by
\be
	S_c(t,z) \propto 
		\exp\left( - {(\tau - \tau_0)^2\over 2 \Delta \tau^2} \right)
	\,\, 	\exp\left( - {(\eta - y_0)^2\over 2 \Delta \eta^2} \right)
\ee
	where the parameters are taken from Table 1, corresponding to the
	best fit to NA44 $S + Pb$ data in this picture.
	Note the relation $(t,z) = (\tau \cosh(\eta), \tau \sinh(\eta) )$.
	The contour-plot of $S_c(t,z)$ on Figure 2 shows that the
	with of particle emission is found to be narrow in proper-time. 
	The parametric plot of $S_c(t,z)$ on Figure 3 indicates the
	long tails of particle production in the $| z| \approx t$ regions,
	the height of the curve is proportional to the production probability.

	Note that in our case the hypothesis that
	pions, kaons and protons are emitted from the same hydrodynamical
	source is in a good agreement with the fitted data. 
	The hypothesis that pion and kaons are emitted from different sources
	was investigated in ref.~\cite{3ds96}, and this hypothesis resulted
	in a worse description of the data than the hypothesis that the
	pions and kaons are produced from the same source.

\subsection{Comparision to hadron - proton reactions at CERN SPS}

	The NA22 collaboration fitted the same model to 250 GeV meson + p
	data~\cite{na22mt}. Their best fit parameter values were
	$\Delta \eta = 1.36 \pm 0.02$, $T_0 = 140 \pm 3$ MeV,
	$\langle u_t \rangle = 0.20 \pm 0.07$ and
	$\langle {\Delta T \over T} \rangle_r = 0.71 \pm 0.14$.
	From a combined HBT and spectrum analysis the
	NA22 experiment finds  a mean freeze-out time of 
	$\tau_0 = 1.4 \pm 0.1$ fm/c,  and a comparable duration
	parameter $\Delta\tau \ge \Delta\tau_* = 1.3 \pm 0.3 $ fm/c.
	The transverse geometrical radius was found to be 
	$R_G = 0.88 \pm 0.13 fm$, which is slightly larger than
	the corresponding Gaussian radius parameter of the proton.
	
	When comparing the source parameters of $S + Pb$ reactions
	to $h + p$ we find that the freeze-out temperature at the
	mean emission time at the center is similar in both cases. 
	The surface temperatures are also equal within errors
	($ 82 \pm 7$ MeV vs. $107 \pm 28 \pm 18 $ MeV). 
	The width of the particle emission in spacetime-rapidity
	$\eta$ is also similar, $ 1.36 \pm 0.02$ vs. 
	$1.6 \pm 0.3 \pm 0.3$. However, we find a significantly
	larger mean freeze-out time in $S + Pb$ reactions
	as compared to $ h + p$ reactions ($5.1 \pm 0.3 \pm 0.3$ fm/c
	vs. $1.4 \pm 0.1$ fm/c). It seems that we observe a 
	sharper freeze-out hypersurface in heavy ion collisions
	than in hadron-proton reactions, cf. $\Delta \tau = 0.3 \pm
	0.3 \pm 0.3$ vs. $\Delta\tau \ge 1.3 \pm 0.3$ fm/c.
	The transversal radius of the matter is also larger
	in $S + Pb $ reactions, $5.4 \pm 0.9 \pm 0.7$ vs.
	$0.88 \pm 0.13$ fm. We also find a much stronger 
	transverse flow in $S + Pb$ than $h + p$ reactions.

\subsection{Comparision to other NA44 data}
	A comparision  of this analysis of NA44 data for
	$S + Pb $ reactions to recent NA44 results~\cite{na44prl} on
	$p + p$, $S + S$ and $Pb + Pb $ reactions  shows that
	the freeze-out temperature in the center at the
	mean freeze-out time, $T_0$, in our case  of $S + Pb$ 
	reactions is within errors similar to the values obtained
	from an analysis of the particle spectra of $p + p$, $S + S$
	and $Pb + Pb$ reactions~\cite{na44prl}. 
	In the $S + Pb$ case, we have a complete description,
	 including the low $p_t$ part of the pion
	spectra with an acceptible $\chi^{2}/ndf$ due to our inclusion 
	of the temperature inhomogeneities. Note also that we do not
	simply reproduce the slopes of particle spectra
	 but  we reproduce the absolute normalized 
	particle spectra for pions,  kaons, as well as the full shape of
	the proton spectra and the HBT radii.

\section{Conclusions}
	A combined fitting of HBT  radius parameters 
	and spectrum fitting is presented
	in the Buda-Lund hydrodynamical parameterization scheme 
	for CERN SPS heavy ion reactions.

	It was necessary to use different particles and absolutely
	normalized particle spectra to obtain a well-defined 
	minimum of each parameters
	for data from the small NA44 acceptance.
	The model is shown to describe
	the data  in a statistically acceptable manner. 
	The parameters can be determined with an accuracy of 10 - 20 \% 
	relative errors including systematic uncertainties.
	The transverse flow and the radial temperature inhomogeneity
	is necessary to achieve this result.

	The same model describes meson - p spectra and correlations at
	CERN SPS
	with a similar central temperature, similar surface temperature
	and similar space-time rapidity width. The main difference between
	$ S + Pb $ and $(\pi/K) + p$ reactions is that the geometrical radii,
	the freeze-out time and the transverse flow are much larger in the first
	case. The particle emission seems to happen
	more suddenly in S + Pb as compared to hadron induced
	reactions at the same energy range. 

\section*{Acknowledgments}
	The authors would like to thank M. Murray and the NA44
	collaboration for making the NA44 data available for us.
        This work was partially supported by the Hungarian NSF 
        grants OTKA - T016206, T024094 and T026435, by the
	Swedish Research Council and by the exchange programme
	of the Hungarian Academy of Sciences and the Royal Swedish
	Academy of Sciences.

\end{document}